\documentclass[twocolumn,showpacs,preprintnumbers,amsmath,amssymb,prl]{revtex4}
\usepackage{amsmath}
\usepackage{amssymb}
\usepackage{revsymb}
\usepackage{graphicx}
\usepackage{dcolumn}
\usepackage{bm}

\begin{document}

\author{A. Thiaville$^1$, Y. Nakatani$^{2,1}$, J. Miltat$^1$ and Y Suzuki$^3$}
\affiliation{$^1$CNRS-Universit\'{e} Paris-sud, Laboratoire de physique des 
solides, B\^{a}t.~510, 91405 Orsay cedex, France, \\
$^2$ University of Electro-communications, Chofu, 182-8585 Tokyo, Japan,\\
$^3$ Osaka University, Dep. of Materials Engineering, Toyonaka, 
560-8531  Osaka, Japan\\}
\pacs{72.25.Ba, 75.70.Kw, 72.25.Pn}
\date{January 21, 2005}

\begin{abstract}
In order to explain recent experiments reporting a motion of magnetic
domain walls (DW) in nanowires carrying a current, we propose a modification
of the spin transfer torque term in the Landau-Lifchitz-Gilbert equation.
We show that it explains, with reasonable parameters, the measured DW
velocities as well as the variation of DW propagation field under
current.
We also introduce coercivity by considering rough wires.
This leads to a finite DW propagation field and finite threshold 
current for DW propagation, hence we conclude that threshold currents 
are extrinsic.
Some possible models that support this new term are discussed.
\end{abstract}
\maketitle
Recent research on magnetic nanostructures has shown that 
effects caused by an electric current flowing across a nanostructure
may dominate over the effects due to the field generated by the same 
current \cite{Slonczewski96,Katine00}.
Most of the work up to now, experimental and theoretical, has been
devoted to the 3-layer geometry (2 magnetic layers separated by a
normal metal spacer, with lateral dimensions well below the
micrometer so as to get single domain behaviour). 
Under current, generation of spin waves \cite{Tsoi98}, layer 
switching \cite{Katine00} and precession of magnetization \cite{Kiselev03}
have been observed.
All these results could be qualitatively explained by the spin transfer model 
\cite{Slonczewski96}.
On the other hand, the situation with an infinite number of layers, namely 
a magnetic nanowire containing a magnetic domain wall (DW), has just started
to be studied. 
Here, under the sole action of a current, the DW may be moved along the 
wire, as confirmed by several experiments
\cite{Klaui03, Grollier03, Vernier04, Yamaguchi04, Lim04, Klaui04}.
The situation is however more complex than with 3 layers, as the 
evolution of the current spin polarization across the DW has to be
described, so that a deep connection with the problem of DW magnetoresistance
exists.
Two limits have been identified by the theories, namely the thin and
thick DW cases.
The length to which the DW width has to be compared is, depending on
the model, the spin diffusion length \cite{Zhang02}, the Larmor 
precession length \cite{Waintal04}, or the Fermi wavelength \cite{Tatara04}.
These lengths are below or of the order of a nanometer in usual 
ferromagnetic metals.
In the experiments cited above (nanowires of typical width 100~nm and 
thickness 10~nm), the DW width is of the order of 100~nm, 
much above these lengths.
Thus, the approximations of current polarization adiabaticity and 
full transfer of angular momentum to the local magnetization apply, and 
the Landau-Lifchitz-Gilbert (LLG) equation of magnetization dynamics becomes
\begin{equation}
\label{eq:LLG1}
\dot{\vec{m}} = \gamma_0 \vec{H} \times \vec{m} + 
\alpha \vec{m} \times \dot{\vec{m}} 
- \left( \vec{u} \cdot \vec{\nabla} \right) \vec{m}.
\end{equation}
The symbols are: $\vec{m}$ unit vector along the local magnetization,
$\gamma_0$ gyromagnetic constant, 
$\vec{H}$ micromagnetic effective field, $M_{\text s}$ spontaneous
magnetization and $\alpha$ Gilbert damping constant.
The velocity $\vec{u}$ is a vector directed along the direction of electron
motion, with an amplitude
\begin{equation}
\label{eq:Defu}
u = J P g \mu_{\text B} / \left( 2 e M_{\text s} \right),
\end{equation}
where $J$ is the current density and $P$ its polarization rate ($u$ is positive
for $P > 0$ {\it i.e.} for carriers polarized along the majority spin
direction).
For permalloy the factor $g \mu_{\text B} /(2 e M_{\text s})$ amounts to
$7 \times 10^{-11}$~m$^3$/C.
The local form (\ref{eq:LLG1}), already introduced by several authors
\cite{Bazaliy98, Ansermet04, Fernandez04, Li04}, can be seen as
the continuous limit of the Slonczewski spin transfer term between
infinitely thin successive cross-sections of the nanowire.

\begin{figure}
\scalebox{0.7}{\includegraphics{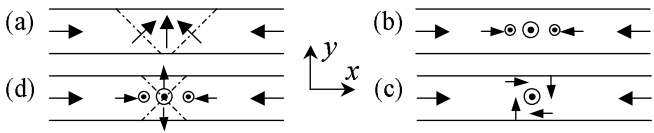}}
\caption{Schematic domain wall structures in a soft nanowire. 
They are governed by magnetostatics and depend on the wire 
cross-section:
(a) transverse wall, (b) perpendicular transverse wall,
(c) vortex wall.
As the fabricated wires are very wide compared to the exchange
length (5~nm for permalloy), the energy maximum situation (b) 
decays to a wall with an antivortex (d) \cite{Nakatani03}.
In cases (a, b) the DW magnetization angle $\phi$ with respect to the 
$y$ axis is 0 and 90$^{\circ}$, respectively.
The stable structures are (a) at smaller and (c) at larger cross-sections
\cite{Nakatani04}.
}
\label{fig:struct}
\end{figure}

The consequences of  (\ref{eq:LLG1}) on the dynamics of a transverse wall
(TW -- Fig.~\ref{fig:struct}a) have been well studied qualitatively 
\cite{Berger86, Tatara04}.
Using the integrated equations for the DW dynamics 
\cite{Slonczewski72, Malozemoff79}, it was found that 2 regimes exist.
At low current, the spin transfer torque is balanced by an internal
restoring torque.
The DW magnetization tilts out of the easy plane (as first
explained by Berger \cite{Berger78}), but
the DW does not move steadily under current.
Above a threshold, the internal torque is not sufficient and DW
motion occurs, together with a continuous precession of the DW
magnetization (in Fig.~\ref{fig:struct}: 
$a \rightarrow d \rightarrow \bar{a} \rightarrow \bar{d}...$, the
bar meaning $\pi$ rotation around the $x$ axis).
The threshold is linked to the steepness of the potential well that
defines the equilibrium orientation of the TW magnetization (mainly
a magnetostatic effect, see Fig.~\ref{fig:struct}).

These qualitative findings were fully confirmed by micromagnetic
computations \cite{Thiaville04} uncovering, however, a
big discrepancy  with experimental results.
The intrinsic current thresholds computed for steady propagation of a TW
in a perfect wire were found to be more than 10 times larger than 
experimental values.
Thermal fluctuations cannot change the situation
because the energy barrier against DW magnetization rotation proves 
very large (for a
$120 \times 5$~nm$^2$ permalloy wire, the energy difference between
states (d) and (a) in Fig.~\ref{fig:struct} is 230~$k_B T_{\text amb}$).
Therefore, we need to introduce corrections to perfect adiabaticity and
pure local spin transfer in the micromagnetic models.
In this letter, we show how the inclusion of a second term describing
the current torque, as introduced previously in the 3 layers
geometry \cite{Heide01}, can resolve this discrepancy.
Our approach is twofold: first we study how micromagnetic results are
modified when this term is phenomenologically included.
In a second step we examine several models of the current effect that
may justify the existence of such a term, and try to estimate its magnitude.

The modified LLG equation, again obtained in the continuous limit,
now reads
\begin{equation}
\label{eq:LLG2}
\dot{\vec{m}} = \gamma_0 \vec{H} \times \vec{m} + 
\alpha \vec{m} \times \dot{\vec{m}} 
- \left( \vec{u} \cdot \vec{\nabla} \right) \vec{m}
+ \beta \vec{m} \times 
\left[ \left( \vec{u} \cdot \vec{\nabla} \right) \vec{m} \right].
\end{equation}
As discussed later, the non-dimensional parameter $\beta$ is much smaller
than unity, hence comparable to $\alpha$.
The solved form of (\ref{eq:LLG2}) shows that the new term modifies
the initial spin transfer torque by a second order quantity, and directly
competes with the damping term associated to the first term.
This is very appealing, as one sees directly from (\ref{eq:LLG1}) that
for zero damping the solution under current is just the solution at zero 
current, but moving at velocity $\vec{u}$.

In order to test the influence of this new term, micromagnetic numerical 
simulations were performed. 
Similarly to previous work \cite{Nakatani03}, a moving calculation region
centered on the DW represented an infinite wire.
Wire width and thickness corresponded to experiments
\cite{Vernier04, Yamaguchi04}, and material parameters 
were those typical of permalloy 
($M_{\text s} = 8 \times 10^5$~A/m, exchange constant $A= 10^{-11}$~J/m
and no anisotropy) with, in most cases, damping fixed to  
$\alpha = 0.02$ \cite{Nakatani03}.
The mesh size was $4 \times 4 \times 5$~nm$^3$.
It was checked first that the Oersted field generated by the current 
had a negligible effect for current values corresponding to experiments.
Indeed, because of the small thickness of the sample, this field is
essentially perpendicular and localized at sample edges.
Its maximum value for an extreme case $u= 100$~m/s, $P= 0.4$ in a
$120 \times 5$~nm$^2$ wire is $\mu_0 H_z = 0.017$~T, a value
well below the perpendicular demagnetizing field (1~T), that
results in a very small out of plane magnetization rotation.
Consequently, all results shown below were computed without the Oersted 
field.
Transverse, but also vortex walls, were considered.

Figure \ref{fig:TW}a shows the wall velocity $v$ under zero field as a
function of the velocity $u$ (proportional to the current) with $\beta$
as a parameter.
The sample is 120~nm wide and 5~nm thick \cite{Vernier04}, with a DW
of the transverse type (Fig.\ref{fig:struct}a).
The $\beta = 0$ curve displays an absence of DW motion for 
$u < u_c = 600$~m/s, and a rapid increase of $v$ towards $u$ above
the threshold, as inferred previously \cite{Tatara04, Thiaville04, Li04}.
Curves with $\beta > 0$ differ markedly from this behaviour, as
DW motion is obtained at any finite $u$ (keep in mind the wire is perfect).
Velocity increases linearly with $u$ and $\beta$, up to a breakpoint where
it decreases (or increases if $\beta < \alpha$ -- not shown).
Results obtained for other $\alpha$ values indicate that all of them can be
described as $v = \beta u / \alpha$ below the breakpoint.
Above breakpoint, the DW structure examination reveals the periodic
injection of antivortices at wire edges, which cross the wire width
and are expelled, similarly to the field driven case \cite{Nakatani03} or to
the current driven case with $\beta = 0$ {\cite{Thiaville04}, 
above their respective thresholds.

The same behavior is obtained at low $u$ for a VW structure in a
larger wire (width 240~nm, thickness 10~nm).
This surprising result (for field-driven propagation, wall mobility 
is much lower for a VW \cite{Nakatani04}) can be explained by the global
arguments of Thiele \cite{Thiele73} about steady-state motions.
Indeed, Eq.~\ref{eq:LLG2} can be transformed to a generalized Thiele
relation
\begin{equation}
\label{eq:Thiele}
\vec{F} + \vec{G} \times \left( \vec{v} - \vec{u} \right)
+ \overleftrightarrow{D} \left( \alpha \vec{v} - \beta \vec{u} \right)
= \vec{0},
\end{equation}
with $\vec{F}$ the static force, $\vec{G}$ the gyrovector (along $z$
here) and $\overleftrightarrow{D}$ the dissipation dyadic \cite{Thiele73}.
The solution is then $\vec{v}= \beta \vec{u} / \alpha$ as long as the
restoring force keeping the vortex in the wire can balance the gyrotropic
term.
Otherwise the VW transforms into a TW by lateral expulsion of
the vortex.
\begin{figure}
\scalebox{0.4}{\includegraphics{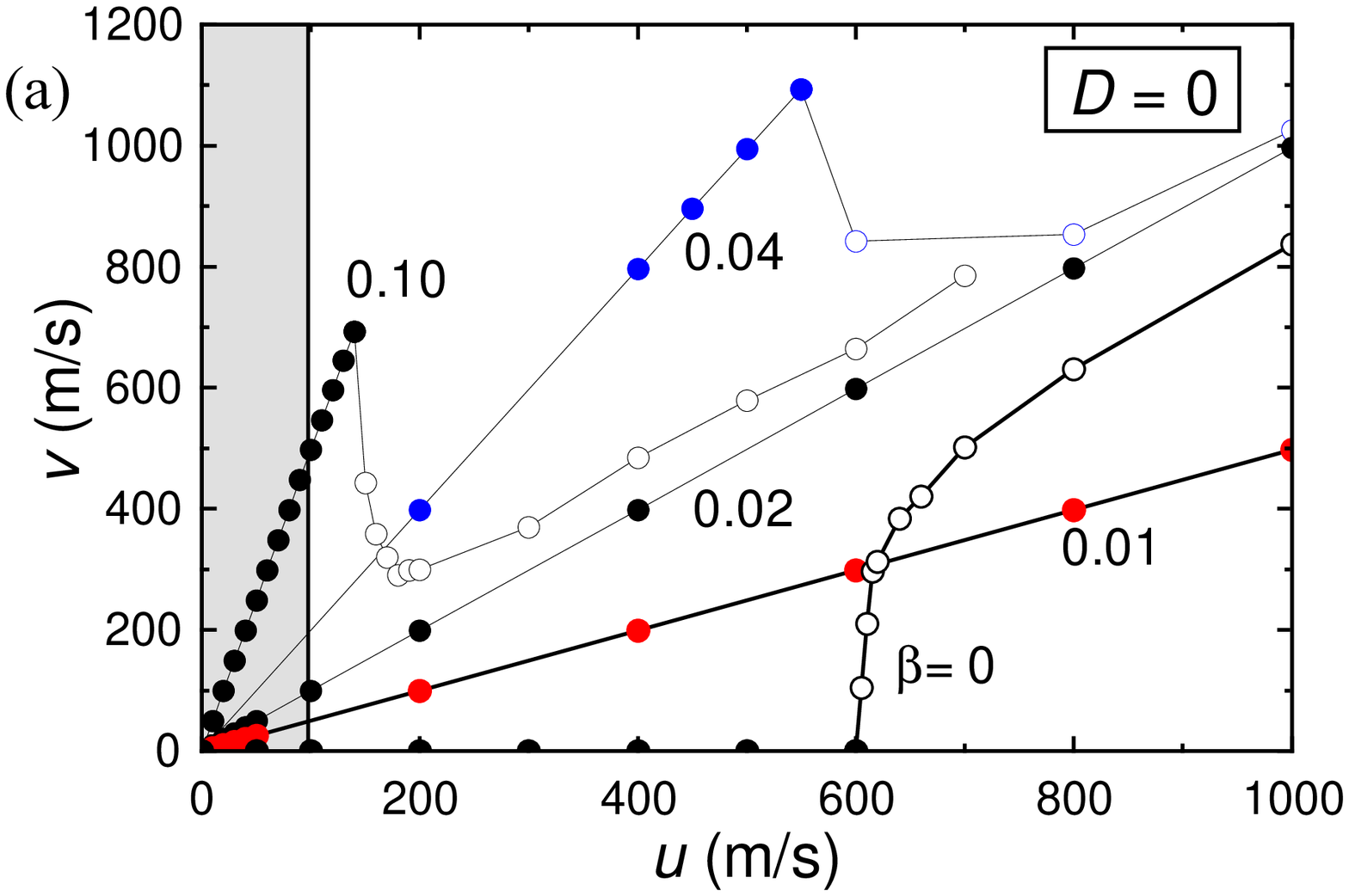}}
\scalebox{0.4}{\includegraphics{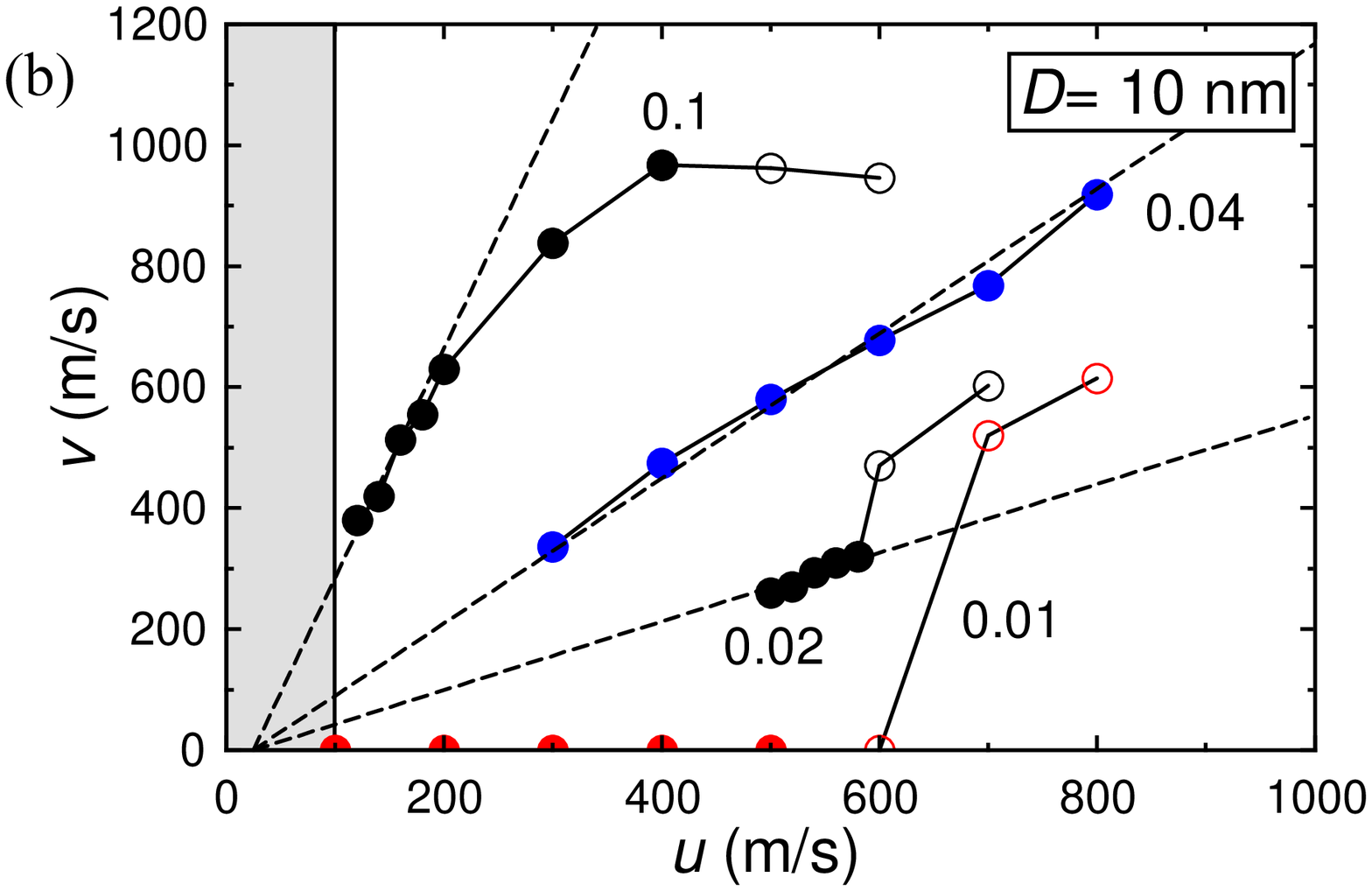}}
\caption{Steady velocity computed for a transverse domain wall by
micromagnetics in a $120 \times 5$~nm$^2$ wire as a function of the
velocity $u$ representing the spin polarized current density
(\ref{eq:Defu}), with the relative weight ($\beta$) of the 
exchange field term as a parameter. 
Open symbols denote vortices nucleation.
The shaded area indicates the available experimental range for $u$.
(a) perfect wire and (b) wire with rough edges 
(mean grain size $D= 10$~nm).
Dashed lines display a fitted linear relation with a 25~m/s offset.
}
\label{fig:TW}
\end{figure}

Experiments on single-layer wires of permalloy 
\cite{Vernier04, Yamaguchi04, Klaui04} show however that a threshold
current exists for DW propagation, similarly to field-driven propagation
(for example, a propagation field $H_{\text P} \sim 20$~Oe was measured in 
Ref.~\cite{Vernier04}).
This is not reproduced by our results for perfect wires.
A roughness of the wire edges was therefore introduced, characterized by a 
mean grain size $D$ (see Ref.~\cite{Nakatani03} for details).
Fig.\ref{fig:TW}b shows how the results of Fig.\ref{fig:TW}a are
modified for $D= 10$~nm, a value which results in an extrapolated
propagation field  $H_{\text P} \sim 10$~Oe, with
velocity vs. field curves given in Ref.~\cite{Thiaville04}. 
At low $\beta$ ($ < 0.02$) the DW is blocked for $u < u_c$, showing that 
roughness-induced DW pinning dominates.
Above $u_c$, the DW moves via nucleation of antivortices and vortices,
again similarly to the perfect case with $\beta = 0$.
If $\beta$ is larger ($0.02 - 0.1$), a window of DW
propagation without vortex nucleation exists (filled symbols),
including values of $u$ well below $u_c$.
Larger $\beta$ values result in DW motion at lower $u$.
However, with our coercivity model and the absence of thermal
fluctuations, continuous DW propagation at ``low'' velocities
($v < 300$~m/s) is not possible as the DW 
eventually stops at positions with a larger wire width fluctuation.
This may be largely due to our coercivity model, as in the same
conditions a VW moves already at $v \approx u= 100$~m/s, and for
regular notches TW propagation at $v < 10$~m/s is obtained.
By comparing with Fig.~\ref{fig:TW}a, it is clear that the rough wire
data cannot be fitted by merely introducing an offset $u_{\text P}$ into the
results for a perfect wire, as done usually in the field-driven case.
On the other hand, simple proportionality of $v$ to $u$ gives 
a very good fit.
However, it would be as good if a small offset existed, as
shown by the dashed lines in Fig.~\ref{fig:TW}b that assume 
$u_{\text P} = 25$~m/s .
Thus, this model of coercivity does create a propagation field $H_{\text P}$ 
and is consistent with a threshold current $u_{\text P}$ for DW propagation, 
even if the data do not allow to determine precisely its value.
An important additional result of the simulations is that, for a rough wire,
the velocity is reduced when compared to the perfect case relation 
$v = (\beta / \alpha) u$.
The reduction factor is not a constant, as
for example at $D = 10$~nm, $\alpha = 0.02$ and $\beta = 0.04$, the factors
for a TW are 0.6, 0.7 and 0.4 for 120x5, 240x5 and 240x10~nm$^2$ strips,
respectively, whereas they are in the range $0.5 - 0.6$ for a VW.

Experimental propagation threshold currents $u_{\text P}$
\cite{Vernier04, Yamaguchi04, Klaui04} for permalloy strips amount to 
$P \times 50$~m/s (TW) and $P \times 80$~m/s (VW), for cross-sections 
120x5, 240x10 and 200x(5-35) nm$^2$, the current polarization $P$ being
unknown.
These values are compatible with our results at $D= 10$~nm, which
was chosen in order to reproduce typical propagation fields.
Although the value of $\beta$ cannot be determined at this stage, we
infer from our results that it should be larger than 0.01.
In addition, the measurement of the propagation field under current
\cite{Vernier04} showed a linear variation of $H_{\text P}$ {\it vs.} $u$, 
with $\Delta H_{\text P}= 3$~Oe for $\Delta u= P \times 100$~m/s. 
As for a perfect wire \cite{notev} one has 
$v= \mu H + \beta u / \alpha$, we
evaluate $\Delta H_{\text P} / \Delta u= \beta /(\alpha \mu)$. 
The mobility calculated by micromagnetics for this nanowire being
$\mu = 26.5$~m~s$^{-1}$Oe$^{-1}$ \cite{Nakatani04}, we estimate, for
$\alpha= 0.02$ and $P=0.4$, $\beta=0.04$.
A micromagnetic calculation for this situation, on the same 120x5~nm$^2$
wire with $\alpha=0.02$, $\beta=0.04$ and $u= \pm 50$~m/s gives
$\Delta H_{\text P} \sim 7.8$~Oe which fully agrees with the above
($0.4 \times 7.8 = 3.1$).
Another experimental data is the DW propagation velocity $v$
(only one measurement, Ref.~\cite{Yamaguchi04}).
It was seen to increase from 2 to 6~m/s as the current velocity was
increased from $P \times 80$ to $P \times 90$~m/s. 
The average slope of these data is consistent with, in a perfect wire,
$\beta = 0.03$, $\alpha= 0.02$, and $P= 0.4$.
Note that (i) the $\beta$ value should be larger because of the 
velocity reduction in a rough wire, and 
(ii) the determination of $\beta$ is subject 
to uncertainty, as the coercivity source we considered may not
be the only meaningful one.
To summarize, we see that the experimental results can be
reproduced by introducing the $\beta$ term in (\ref{eq:LLG2}), with a value
of about 0.04 for permalloy.

\begin{figure}
\scalebox{0.4}{\includegraphics{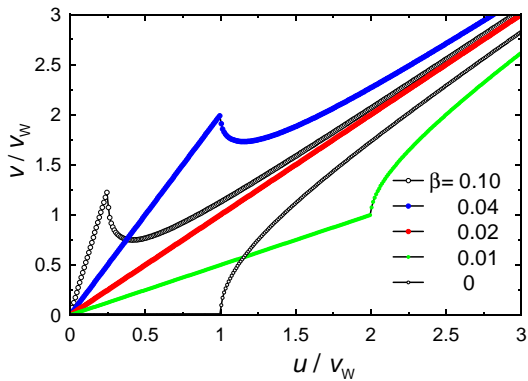}}
\caption{Domain wall propagation velocity under current as obtained
from the one-dimensional model, assuming a perfect wire and 
a constant wall
width parameter $\Delta$, for a damping constant $\alpha= 0.02$.
The curves' shape is extremely similar to the numerical results
(Fig.~\ref{fig:TW}a).
Velocities are normalized to the Walker threshold
$v_{\text W} = \gamma_0 \Delta H_{\text K} / 2$.}
\label{fig:ana}
\end{figure}
We describe now how the numerical micromagnetic results shown above 
can be qualitatively understood in the TW case, in
the framework of the extended Slonczewski equations describing DW motion.
These are constructed by assuming a given DW profile with some parameters
(here the wall position $q$ and magnetization angle 
$\phi$ -- see Fig.\ref{fig:struct}).
In the case of pure spin transfer the LLG equation could be
derived from a least action principle, allowing an easy
derivation of the equations \cite{Thiaville04}. 
This is no longer true with the additional term and the original 
procedure must be applied
\cite{Slonczewski72, Malozemoff79}.
For a one-dimensional TW profile, characterized
by a DW width parameter $\Delta$, one then gets
\begin{equation}
\label{eq:Sloncphi}
\dot{\phi} + \alpha \dot{q} / \Delta =
\gamma_0 H_{\text a} + \beta u / \Delta 
\ \ ; \ \ 
\dot{q} / \Delta - \alpha \dot{\phi} =
\gamma_0 H_{\text K} \sin \phi \cos \phi + u / \Delta,
\end{equation}
with $\Delta = \Delta(\phi)$ the wall width parameter minimizing 
the DW energy at wall magnetization angle $\phi$.
In the simplest 1D model \cite{Thiaville04} one has
$\Delta(\phi) = \left( A / \left( K_0 + K \sin^2 \phi \right) 
\right) ^{1/2}$, with $K_0$ the effective longitudinal anistropy
and $K$ the transverse one 
($H_{\text K}= 2 K / \left( \mu_0 M_{\text s} \right)$).
One sees from (\ref{eq:Sloncphi}) that the additional term enters the
equations similarly to an applied field
$H_{\text equiv}= u \beta / (\gamma_0 \Delta)$,
explaining DW motion at any $u$ in a perfect wire.
At zero applied field one gets
\begin{equation}
\dot{\phi} = \frac{\alpha}{1+\alpha^2} \left[
\frac{\beta - \alpha}{\alpha} \frac{u}{\Delta}
-\frac{\gamma_0 H_{\text K}}{2} \sin 2\phi \right]
\ \ ; \ \ 
\dot{q} = \beta u / \alpha - \Delta \dot{\phi} / \alpha.
\label{eq:solana}
\end{equation}
For $|u| < \left( \gamma_0 H_{\text K} / 2 \right) \Delta(\pi / 4)
\alpha / |\beta - \alpha |$ \cite{noteD}
a stationary regime exists with
$\dot{q} = \beta u / \alpha$.
The DW velocity is remarkably independent on the DW width, as
already shown with (\ref{eq:Thiele}).
Above the threshold $\phi$ oscillates and so does the velocity.
The average wall velocity, plotted in Fig.\ref{fig:ana}, is
very similar in behaviour to numerical results (Fig.\ref{fig:TW}a).
The non-trivial effect of roughness is not included in this 1D model,
but if one takes it into account by an increased dissipation (larger
$\alpha$) a velocity reduction in rough wires is natural.

We finally turn to discussing possible origins of 
Eq.\ref{eq:LLG2}.
One may first remark that the new term exhausts the possibilities 
of introducing linearly the magnetization gradient along the 
electric current direction into the LLG equation.

The approach of Tatara and Kohno \cite{Tatara04} put forward two
effects of an electric current across a DW, called spin transfer 
and momentum transfer.
The former, dominant for thick DWs, has the same form as our first
term while the latter, dominant for thin walls, enters the integrated
DW dynamics equations (\ref{eq:solana}) similarly to 
our second term.
Thus, for a DW of finite thickness the second term should remain present
in some proportion. 
As the characteristic length is here the Fermi wavelength \cite{Tatara04},
we expect $\beta \ll 1$. Note however the lack of an expression in the 
intermediate DW thickness regime that would predict the value of $\beta$.

Zhang {\it et al.} calculated spin accumulation for a current flowing
across a multilayer, and the resultant torque on the magnetization, 
with the inclusion of the {\it s-d} exchange interaction 
\cite{Zhang02,noteW}.
Extending their model \cite{Zhang02} to the case of a linearly 
varying magnetization
gives both terms of (\ref{eq:LLG2}) with a slightly modified equivalent
velocity $u$ and a $\beta$ factor given by
\begin{equation}
\label{eq:betaYS}
\beta = \left( \lambda_J / \lambda_{\text sf} \right) ^2
= \hbar / \left( J \tau_{\text sf} \right) ,
\end{equation}
with $J$ the {\it s-d} exchange interaction energy, $\tau_{\text sf}$ the
spin-flip time, $\lambda_J$ and $\lambda_{\text sf}$ the associated
diffusion lengths.
Note that with $\lambda_J = 1$~nm and $\lambda_{sf} = 5$~nm for permalloy,
one gets $\beta = 0.04$.
As for pure metals $\lambda_{\text sf}$ can be much larger
\cite{Barthelemy99}, one expects
that $\beta$ may become smaller and, hence, that DWs may
prove more difficult to move.
This should be investigated by experiments.
One should also be aware that the time $\tau_{\text sf}$ in
(\ref{eq:betaYS}) may not be exactly the spin-flip time. 
This time appears in fact as the transverse relaxation time in 
the evolution equation of the spin accumulation. 
If part of this relaxation occurs through interaction with the 
localized moments, for example by incoherent precession around the
local magnetization, then the moment is transferred to the localized
spin system and does not contribute to $\beta$.

In conclusion, we have shown that the inclusion of a second term
describing the spin transfer torque, akin to the exchange field term
considered previously, might solve the puzzle of the
quantitative explanation by micromagnetics of the current induced DW
motion \cite{noteZ}.
This view is supported by a direct comparison of experimental results
(DW velocity, reduction of DW propagation field under current) to
micromagnetic simulations of realistic nanowires (excepting pinning 
phenomena).
The model predicts that DWs should move under smaller currents for
more perfect wires (having a lower propagation field).
As our approach was phenomenological, a basic understanding of the form 
of the spin-transfer torque in continuous structures as wide as a domain 
wall is still very much needed.
The magnitude of the new term, and its material dependence, should allow 
to discriminate between models.
\acknowledgments
A.T. acknowledges fruitful discussions with F. Pi\'{e}chon and N. Vernier.


\begin{thebibliography}{99}

\bibitem{Slonczewski96}
J.~C.~Slonczewski,
J.~Magn.~Magn.~Mater. {\bf 159}, L1 (1996).

\bibitem{Katine00}
J.~A.~Katine {\it et al.},
Phys.~Rev.~Lett. {\bf 84}, 3149 (2000).

\bibitem{Tsoi98}
M.~Tsoi {\it et al.},
Phys.~Rev.~Lett. {\bf 80}, 4281 (1998);
{\bf 81}, 493(E) (1998).

\bibitem{Kiselev03}
S.~Kiselev {\it et al.},
Nature {\bf 425}, 380 (2003).

\bibitem{Klaui03}
M.~Kl\"{a}ui {\it et al.},
Appl.~Phys.~Lett. {\bf 83}, 105 (2003).

\bibitem{Grollier03}
J.~Grollier {\it et al.},
Appl.~Phys.~Lett. {\bf 83}, 509 (2003).

\bibitem{Vernier04} 
N.~Vernier {\it et al.}, 
Europhys.~Lett. {\bf 65}, 526 (2004).

\bibitem{Yamaguchi04}
Y.~Yamaguchi {\it et al.},
Phys.~Rev.~Lett. {\bf 92}, 077205 (2004).

\bibitem{Lim04}
C.~K.~Lim {\it et al.},
Appl.~Phys.~Lett. {\bf 84}, 2820-2822 (2004).

\bibitem{Klaui04}
M.~Kl\"{a}ui, U.~R\"{u}diger, C.A.F.~Vaz, J.A.C.~Bland, W.~Wernsdorfer, 
L.J.~Heyderman, F.~Nolting, G.~Faini, and E.~Cambril,
Phys. Rev. Lett., accepted.

\bibitem{Zhang02}
S.~Zhang, P.~M.~Levy and A.~Fert,
Phys.~Rev.~Lett. {\bf 88}, 236601 (2002).

\bibitem{Waintal04} 
X.~Waintal and M.~Viret, 
Europhys.~Lett. {\bf 65}, 427 (2004).

\bibitem{Tatara04}
G.~Tatara and H.~Kohno,
Phys.~Rev.~Lett. {\bf 92}, 086601 (2004).

\bibitem{Berger86}
L.~Berger,
Phys.~Rev.~B {\bf 33}, 1572 (1986).

\bibitem{Slonczewski72}
J.~C.~Slonczewski,
Int.~J.~Magnetism {\bf 2}, 85 (1972).

\bibitem{Malozemoff79}
A.~P.~Malozemoff and J.~C.~Slonczewski,
{\it Magnetic domain walls in bubble materials}
(Academic Press, New York, 1979).

\bibitem{Berger78}
L.~Berger,
J.~Appl.~Phys. {\bf 49}, 2156-2161 (1978).

\bibitem{Thiaville04}
A.~Thiaville {\it et al.},
J.~Appl.~Phys. {\bf 95}, 7049-7051 (2004).

\bibitem{Bazaliy98}
Ya.~B.~Bazaliy, B.~A.~Jones and S-C.~Zhang,
Phys.~Rev.~B {\bf 57}, R3213 (1998).

\bibitem{Ansermet04}
J-P.~Ansermet,
IEEE~Trans.~Magn. {\bf 40}, 358 (2004).

\bibitem{Fernandez04}
J.~Fern{\' a}ndez-Rossier {\it et al.},
Phys. Rev. B {\bf 69}, 174412 (2004).

\bibitem{Li04}
Z.~Li and S.~Zhang,
Phys.~Rev.~Lett. {\bf 92}, 207203 (2004).

\bibitem{Heide01}
C.~Heide, P.E.~Zilberman and R.J.~Eliott,
Phys.~Rev.~B {\bf 63},064424 (2001).

\bibitem{Nakatani03}
Y.~Nakatani, A.~Thiaville and J.~Miltat,
Nature Mater. {\bf 2}, 521 (2003).

\bibitem{Nakatani04}
Y.~Nakatani, A.~Thiaville and J.~Miltat,
J. Magn. Magn. Mater., to appear.

\bibitem{Thiele73}
A.~A.~Thiele,
Phys.~Rev.~Lett. {\bf 30}, 230 (1973).

\bibitem{notev}
This just adds the effects of field and current. The 1D model
(\ref{eq:Sloncphi}) derives this relation directly under the
hypothesis of stationarity.

\bibitem{noteD}
Strictly speaking, the maximum is not obtained at $\phi = \pi/4$;
it depends on the ratio of effective transverse to longitudinal
anisotropy $K / K_0$. The relation given is an underestimate by less 
than 10\% for $K / K_0 < 10$.

\bibitem{noteW}
The approach of Ref.\cite{Waintal04} starts from the same
hamiltonian, but solves a Schr\"odinger equation instead of the
diffusion equation with relxation term considered here.
The result is a $\beta$ term which oscillates in space. 

\bibitem{Barthelemy99}
A.~Barth\'el\'emy, A.~Fert and F.~Petroff,
in {\it Handbook of Magnetic Materials, vol.12}, K.H.J. Bushow Ed.,
(Elsevier, Amsterdam, 1999).

\bibitem{noteZ}
Just before submitting this paper, we discovered the independent work 
of Zhang and Li (cond-mat/0407174, Phys. Rev. Lett., 
{\bf 93} 127204 (2004))
proposing a derivation of the $\beta$ term
and an investigation of some of its consequences.

\end{thebibliography}
\end{document}